# Regional Development in the Knowledge-Based Economy: The Construction of Advantage

Phil Cooke* & Loet Leydesdorff**

* Centre for Advanced Studies, Cardiff University, Cardiff CF10 3BB, Wales (UK); cookePN@cardiff.ac.uk; http://www.cf.ac.uk/cass
** Amsterdam School of Communications Research (ASCoR), University of Amsterdam, Kloveniersburgwal 48, 1012 CX Amsterdam, The Netherlands; loet@leydesdorff.net; http://www.leydesdorff.net

**Abstract**
In this introduction the editors showcase the papers by way of a structured project and seek to clarify the two key concepts cited in the title. We consider the history of the idea that knowledge is an economic factor, and discuss the question of whether regions provide the relevant system of reference for knowledge-based economic development. Current transformations in university-industry-government relations at various levels can be considered as a metamorphosis in industry organization. The concept of *constructed advantage* will be elaborated. The various papers arising from a conference on this subject hosted by Memorial University, Newfoundland, Canada are approached from this perspective.

**JEL Classification; A14, O33, R11, R58**

## Introduction

If we are to make progress in understanding the transformations occurring in economic relations today, it is important to clarify key elements of interest and the perspectives from which they are being observed. Thus, we shall reflect the ethos of the papers that follow, and the conference that gave rise to them, by highlighting and defining two key terms in full knowledge that important elements of both are contested rather than settled descriptions of reality.

*Knowledge economy* and *knowledge-based economy* are common terms nowadays that are often used synonymously. However, this does not settle the question of whether or not the two expressions actually mean the same thing. We shall argue that 'knowledge economy' is the older of the two concepts, with its origins in the 1950s. It focused mainly on the composition of the labour force. The term 'knowledge-based economy' has added the structural aspects of technological trajectories and regimes from a systems perspective. This perspective leads, for example, to discussions about intellectual property rights as another form of capital.

The regional dimension of analysis and policy for enhanced economic development provides another contested area with the notion of 'regional development', but in particular the 'regional' element of this couplet. A 'regional innovation system' combines the focus on regions with a systems perspective. On the occasion of a previous issue focusing on European regions, we have argued that the trajectory of a region can be the subject of evolution when systemic innovations are involved (Leydesdorff, Cooke & Olazaran, 2002).

The term 'innovation' as widely used in economics and related sub-disciplines is also broad and variable. Hence, if confronted with a definition like 'the commercialisation of new knowledge,' a practising innovator is likely to wish to debate this meaning. 'Region' has some of these characteristics, too. Thus geopolitics has appropriated it to denote subcontinental, geographically neighbouring areas of the globe like the 'Middle East', 'Balkan' or 'Baltic' regions, rather as the term 'theatre' was appropriated for a smaller area in which wars are conducted. It would be more accurate to call the former 'georegions' to distinguish neighbouring segments of the world from the dictionary definition of 'region'.

The following papers arose from a conference hosted from 3-5 October 2003 by Memorial University, St. John's, Newfoundland, Canada with the title 'Knowledge-Based Economy and Regional Economic Development: An International Perspective'. Among the sponsors were the UK Regional Studies Association, represented by chief executive Sally Hardy, and numerous regional development agencies representing the Maritime Provinces of Canada. The latter, particularly, had a lively interest in the implications for development in their regions, represented by Provincial Administrations like the Government of Newfoundland & Labrador, the Atlantic Canada Opportunities Agency, and the Atlantic Provinces Economic Council. Similarly, sponsorship by Industry Canada, and Human Resources Development Canada showed the federal level to be a concerned policy actor keen to raise its absorptive capacity about the regional implications and effects of the rising knowledge economy. The conference was successfully organized by Dr. Wade Locke and Prof. Scott Lynch of the Department of Economics, Memorial University, and this Special Issue of the *Journal of Technology Transfer* is dedicated to them with thanks.

**Regional Development**

Region has its origin in the Latin *regio* which stems from *regere*, meaning 'to govern'. In the field of regional development, 'region' has been used precisely in this sense, namely to signify the governance of policies to assist processes of economic development. Here, the concept of 'region' as administratively defined is of primary importance. The administrative dimension leads to the following definition: *region is an administrative division of a country*. Thus, for example, 'Tuscany is a region of Italy'.

Of course, there are other definitions. A region can also be 'any large, indefinite and continuous part of a surface or space', or 'a unit for geographical, functional, social or cultural reasons', or in military usage 'the part of the theatre of war not included in the theatre of operations'. Thus, 'region' is considered as an abstract space, a cultural area, or a military field of action. None of these definitions captures the precision required and supplied by the administrative definition.

'Regional' is nested territorially beneath the level of the country, but above the local or municipal level. In objective terms, this is generally how the conceptual level can be aligned with the geographical one. However, some countries only have national states and local administrations, but no regions. Some of these, like Finland and Sweden, are evolving toward regional administrations. But can they then be said to experience 'regional development'? We submit that they can; and by dint of national or even supranational policy for regional development, or local proactivity, possibly including local collaborative partnerships of municipalities pursuing aims of *constructed advantage*, they often do.

**The Knowledge Economy**

It may surprise some readers that using the word 'knowledge' as a structural component from an economic perspective is not a new idea. Apart from Marx, who exempted mathematics and the natural sciences from the direct influence of the social and economic infrastructure, and argued that superstructures were not only mere reflections of infrastructures but could in turn react upon them (see Coser, 1977), it was Schumpeter who first recognized the importance of knowledge in the economy by his reference to 'new combinations of knowledge' at the heart of innovation and entrepreneurship (Schumpeter, 1911, p. 57). Nonaka & Takeuchi (1995) also show that Marshall (1916) recognized that:
> 'Capital consists in a great part of knowledge and organisation... knowledge is our most powerful engine of production... organisation aids knowledge' (p. 115)

Typically, however, neoclassical economics neglected what was not contained in price information and made no effort to add to economic knowledge by trying to measure its economic contribution. Thereafter, Hayek (1945, 1948) identified 'the division of knowledge as the really central problem of economics as a social science' (1948, p. 51) and saw its key question as the puzzle of how localized knowledge held by fragmentary firms and individuals nevertheless produces an ordered market demand and supply:
> 'The most significant fact about this system is the economy of knowledge with which it operates, or how little the individual participants need to know in order to be able to take the right action. In abbreviated form, by a kind of symbol, only the most essential information is passed on, and passed on only to those concerned' (Hayek, 1948, p. 86)

None of these writers was writing about the knowledge economy *per se* but rather its fundamental importance to the functioning of all aspects of the economy from innovation to production, organization, and markets.

A further progenitor of the view that knowledge is a most important economic resource was Penrose (1959). She founded what has now evolved into the 'dynamic capabilities of firms' approach to microeconomics (Teece & Pisano, 1996). She referenced the firm's characteristics as an administrative organization (after Marshall, 1916 and Coase, 1937) and home to accumulated human and material resources. The latter are inputs to services rendered, and these are the product of the firm's accumulated knowledge:
> 'a firm's rate of growth is limited by the growth of knowledge within it, but a firm's size by the extent [of] administrative efficiency' (Penrose, 1995, xvi-xvii)

In effect, in the words of Nonaka & Takeuchi (1995), 'the firm is a repository of knowledge'(p. 34). Penrose (1995) also acknowledged that had the term been available in the 1950s, she would have referred to the dynamic capabilities of firms residing in *knowledge networks* (Quéré, 2003). Thus, Penrose (1995) noted the following crucial feature of the massively increased value of transferable knowledge to the wider economy for the firm:
> 'the rapid and intricate evolution of modern technology often makes it necessary for firms in related areas around the world to be closely in touch with developments in the research and innovation of firms in many centres' (Penrose, 1995, p. xix)

Importantly, Penrose continues, the rise of business knowledge networks represents a *metamorphosis* in the contemporary economy. The key to the knowledge-based economy is at least partly revealed as this metamorphosis in the nature of industry organization to facilitate interaction with valuable knowledge, and not to conceal it, as was common in the previous phase of the global economy.

**The Knowledge-based Economy**

Whereas the concept of a 'knowledge economy' emerged within the context of the economic analysis of the quality of the input factors in the production process (Schumpeter, 1939), the term 'knowledge-based economy' finds its roots in more recent discussions from a systems perspective (e.g., Sahal, 1981, 1985). National governments, for example, need a systems perspective for developing science, technology, and innovation policies (Nelson, 1982). Benoît Godin discusses the various definitions of a knowledge-based economy in his contribution to this issue entitled 'The Knowledge-Based Economy: Conceptual Framework or Buzzword?'

By the second half of the 1950s, it had become increasingly clear to both policy makers and economic analysts that the continuing growth rates of Western economies could no longer be explained in terms of traditional economic factors such as land, labour, and capital. The 'residue' (Abramowitz, 1956; OECD, 1964) had to be explained in terms of the upgrading of the labour force, surplus generated by interaction effects, and more generally the role of knowledge in the economy (Rosenberg, 1976). The *Organization for Economic Co-operation and Development (OECD)* was created in 1961 in order to organize and to coordinate science and technology policies among its member states, that is, the advanced industrial nations.[1] This led in 1963 to the *Frascati Manual* in which parameters were defined for the statistical monitoring of science and technology on a comparative basis.

It is but a short step to link insights like these to the earliest work to operationalize a notion of the 'knowledge economy' arising from the pioneering work conducted by Machlup (1962). He sought to identify those sectors with a heavy concentration of knowledge assets. He next attempted to map the production and distribution of knowledge sectors in the United States economy. Machlup classified knowledge production into six major sectors: education, R&D, artistic creation, communications media, information services, and information technologies. He showed that these account for the largest sectoral share of GDP and employment in the economy, and predicted that this share was destined to grow both absolutely and relatively over time. With brief interventions from Eliasson *et al.* (1990) and Burton-Jones (1999) who further specified the knowledge intensity of sectors by value and labour qualifications respectively, we reach the statements of the Organization for Economic Cooperation and Development (1996, 1999) calling for the measurement of the knowledge-intensity of national and regional economies (OECD/Eurostat, 1997).

Studies of the knowledge-based economy focus not only on human capital, but also on the sectoral characteristics of the knowledge factor (Nelson, 1982; Pavitt, 1984). Technological trajectories and regimes shape innovation systems, but with a dynamics different from those of economic or geographical factors (Nelson & Winter, 1982). The recombination of the economic dynamics of the market, the dynamics of knowledge-based innovation, and governance generates the systems perspective. An innovation system can then be defined at the national level (Freeman, 1987, 1988; Lundvall, 1988, 1992; Nelson, 1993), at the regional level (Cooke, 1992; Cooke *et al.*, 2004), or in terms of a dynamic model like the Triple Helix of university-industry-government relations (Etzkowitz & Leydesdorff, 2000; Leydesdorff, 1994).

---

[1] The OECD was based on the OEEC, the Organization for European Economic Cooperation, that is, the organization which had served for the distribution of the U.S. and Canadian aid under the Marshall Plan during the postwar period.

A key paper in this collection by Dafna Schwartz entitled 'The Regional Location of Knowledge-Based Economy Activities in Israel' uses longitudinal regional data made available by the Israeli Statistical Service. Schwartz shows empirically what Myrdal (1957) and Hirschman (1958) had theorized decades earlier. Myrdal (1957) proposed that spatial development is characterized by *'cumulative causation'* with associated *'spread'* and *'backwash'* effects. This implies Krugman's (1995) increasing returns to scale (through 'backwash') and developmental 'spread' to other nearby areas. Hirschman's (1958) elaboration on this was that 'spread' would be driven by the innovative capacity of competing technology users. Under the conditions of a knowledge-based economy, the key spatial hypothesis is that, over relatively short time periods, core cities grow through increasing returns (to knowledge), with the 'satellites' of leading technology innovators 'spread' by knowledge exploitation or commercialization nearby.

Static pictures of the UK and EU have been consistent with this expectation (Cooke, 2002; Cooke & De Laurentis, 2002). However, Schwartz's dynamic picture of spatial divergence in Israel 1995-2002 is consistent with Myrdal and Hirschman rather than Krugman, who modeled spatial increasing returns under conditions of imperfect knowledge as a zero-sum game resulting in a single spatial monopoly.

More recently, Krugman (2000) himself warned that his 'two-locations competing' models can be misleadingly 'simplistic'. However, Krugman's modelling seems to gain support in the work presented in this collection by Carol Robbins entitled 'The Impact of Gravity-Weighted Knowledge Spillovers on Productivity in Manufacturing.' Robbins uses U.S. data to show increasing knowledge returns to scale from localized knowledge spillovers in six key industries. Scott Tiffin and Gonzalo Jimenez, in their contribution entitled 'Design and Test of an Index to Measure the Capability of Cities in Latin America to Create Knowledge-Based Enterprises,' develop a measurement instrument in order to reveal disparities in Latin America. The message also aligns with already published research reported at the Memorial University conference showing three Canadian metropolitan areas to be aggregating an overwhelming amount of knowledge-economy sectoral assets while medium cities and the periphery are losing theirs (Polèse, 2002).

**The Impact of Regions**

The general argument about the salience of the organization of knowledge in the sectoral, skills, and spatial composition of the economy embraces the position of Castells (1996), who is widely known for the observation that productivity and competitiveness are, by and large, a function of knowledge generation and information processing, and that this has involved a Penrose-type metamorphosis entailing a different mode of thinking about economies. Thus the balance between knowledge and resources has shifted so far towards the former that knowledge has become by far the most important factor determining standards of living— more important than land, capital, or labour. Today's most advanced economies are fundamentally knowledge-based (Dunning, 2000). Even neoclassicists like Paul Romer recognize that technology (and the knowledge on which it is based) has to be viewed as an equivalent third factor along with capital and land in leading economies (Romer, 1990). Inevitably this leads to issues of the generation and exploitation of knowledge. How is the system of knowledge production organized and controlled? (Whitley, 1984, 2001; Leydesdorff, 1995).

In a knowledge-based economy, inequality is generated by mechanisms of inclusion and exclusion only partially overlapping those of a traditional (capitalist) economy. With less emphasis, one can also say that another variant of capitalism is induced (Hall & Soskice, 2001). The mechanisms of inclusion and exclusion are no longer tightly coupled to one's class position in the production process as in an industrial economy. The geographical component can be expected to play an independent role in the knowledge-based dynamics because the newly emerging system is grounded in communication networks.

Burton-Jones (1999) noted that the gap between rich and poor nations is accelerating under 'knowledge capitalism.' Knowledge-intensity can also lead to a growing gap within societies. Knowledge-intensive dynamics of scale and scope induce mechanisms for the retention of wealth that are different from the dynamics of mass production. The increasing role of the service sector, notably, generates another dynamic (Barras, 1990). In this collection, Carla de Laurentis discusses 'Digital Knowledge Exploitation: ICT, Memory Institutions, and Innovation from Cultural Assets' as an example of this process.

The work that has been done spatially to map the knowledge-based economy shows how disequilibriating its effects can be. The core city moves away statistically from the periphery, as in Canada and elsewhere, in the intensity with which it accumulates knowledge-based activities. Simultaneously, new high technology satellite towns 'swarm,' to use a Schumpeterian term, around the mother city. Even static analysis reveals this pattern, with some satellites scoring much higher than the main city around which they aggregate. Peripheral islands and regions or localities may score as low as 37% of the index average of 100% compared to 157% for Stockholm (e.g. Aegean Islands in the EU context; Cooke & De Laurentis, 2002; Dannell & Persson, 2003). Compared to GDP disparities a five-to-one ratio in the knowledge economy measure is approximately twice that given by measuring economic welfare differences more conventionally.

Hence, for the industries of the future, the core cities are highly privileged in most countries while the peripheries are generally impoverished and becoming more so, presaging major out-migration of youth and the metamorphosis of such areas into socially deserted or playground economies. The policy imperative to devise mechanisms by which non-metropolitan regions may, in future, participate in the knowledge-based economy is clearly overwhelming. For example, Godfrey Baldacchino's paper in this collection entitled 'Small Islands versus Big Cities: Lessons in the Political Economy of Regional Development from the World's Small Islands' points to a competitive advantage enjoyed by some islands.

**Constructed Advantage**

It is time to say more about this term and to offer what we say as a context for other papers that follow this Introduction. It has been suggested that the idea originates with Adam Smith, but Foray & Freeman (1993) re-introduced it yet scarcely explored it. More attention has been devoted to it in comparison to other well-known forms of economic advantage by De la Mothe & Mallory (2003), as follows:

- *Comparative Advantage* - Regions have been a focus for economists who viewed them through the lens of development economics usually set in a framework of comparative advantage. This idea, deriving from David Ricardo and trade theory, explained economic welfare in terms of initial resource endowments traded between regions and nations. Thus, cotton goods enjoying a comparative production advantage from mercantile and climatic conditions in northwest England were traded with Port wine from Portugal's Norte region enjoying a comparable mercantile and climatic comparative advantage. While policies were not excluded from such an analysis, they mainly added up to forms of mercantilism, and Ricardo advocated intervention regarding technological change. The overwhelming framework which government policy gave rise to and which promoted comparative advantage was *laissez-faire.*

- *Competitive Advantage* - By the mid-1970s, visible cracks were appearing in the economic models and frameworks that characterize pure comparative advantage. Thus countries with a large labour supply would naturally export goods that were labour-intensive (e.g., China), while countries that were technologically advantaged (e.g., the United States) produced and exported technologically advanced products. The paradox arose when advanced economies exported labour-intensive goods as well as technologically intensive goods. The key weakness was the failure to acknowledge technological *process* change as well as product innovation as being endogenous to economic growth. Krugman (1995) and Porter (1990, 1998) noted the *competitive* advantage of firms in which distributed supply chains and the role of large domestic markets became accepted, and saw this advantage as central to explanations of inter-firm and firm-market success. Intra-industry trade and localized demand conditions for market competitiveness were highlighted. But no explanation was offered for why some regions prosper while others do not. The emphasis on markets meant that funding and policy support by the public sector was largely ignored.

- *Constructed Advantage* - The analytic observations of the two preceding perspectives do not embrace the new dynamics of innovation and the capacity to exploit them which are essential to growth. The 'new competitive advantage' (Best, 2001) highlights regional development economics, the dynamic of which draws upon *constructed advantage.* This knowledge-based construction requires interfacing developments in various directions:
    - *Economy* – regionalization of economic development; 'open systems' inter-firm interactions; integration of knowledge generation and commercialization; smart infrastructures; strong local and global business networks.
    - *Governance* – multi-level governance of associational and stakeholder interests; strong policy-support for innovators; enhanced budgets for research; vision-led policy leadership; global positioning of local assets.
    - *Knowledge Infrastructure* – universities, public sector research, mediating agencies, professional consultancy, etc. have to be actively involved as structural puzzle-solving capacities.
    - *Community and culture* – cosmopolitanism; sustainability; talented human capital; creative cultural environments; social tolerance. This public factor provides a background for the dynamics in a Triple Helix of university-industry-government relations (Leydesdorff & Etzkowitz, 2003).

Hence, *constructed advantage* is both a means of understanding the noted metamorphosis in economic growth activity and a strategic policy perspective of practical use to business firms, associations, academics, and policy makers.

In the Triple Helix model constructed advantages have been conceptualized as the surplus value of an overlay of relations among the three components of a knowledge-based economy: (1) the knowledge-producing sector (science), (2) the market, and (3) governments. Those places with research universities witness a growing demand for knowledge transfer to industry and, through government, to society (Etzkowitz & Leydesdorff, 1998; Etzkowitz *et al*., 2000). Moreover, the geographical spread of universities is reasonably uniform in advanced industrial countries. For research knowledge, industry and government can be expected to pay more for privileged access to knowledge-based growth opportunities by funding research, stimulating closer interactions among the three institutional partners, subsidizing infrastructure (e.g., incubators and science parks), and stimulating academic entrepreneurship skills and funding.

The exemplar *par excellence* of this phenomenon has been the Massachusetts Institute of Technology (Etzkowitz, 2002). In this issue, Chrys Gunasekara's paper entitled 'Reframing the Role of Universities in the Development of Regional Innovation Systems' uses the Triple Helix model to investigate non-metropolitan universities in the quite different context of Australia. Not surprisingly, he finds that a model design based on MIT works poorly for the more typical universities and regions that serve as his laboratory. Nevertheless, the principles of Triple Helix *rapprochement* are valid also for such distinct 'epistemic communities' (Haas, 1992) as the three implied, but the boundary-crossing effort required can defeat the unwary.

Early work on regional innovation systems (Cooke, 1992; Cooke & Morgan, 1994) attempted to capture the integrative and interactive nature of the knowledge-based economy examined from the regional perspective. The list of networking partners includes the base institutions like universities, research laboratories, research associations, industry associations, training agencies, technology transfer organizations (TTOs), specialist consultancies, government development, technology and innovation advisory agency programme-funding, and private investors. This knowledge exploration, examination and exploitation base supports the innovation efforts of large and small firms in many industries. Not all interactions are only intra-regional; many are also national and global, but in the most accomplished regional economies like Baden-Württemberg, a majority of such institutional networking interactions were regional, and on such regular terms that the networking had become systemic (Cooke, 2001).

The variability achieving such seamless interaction is focused on incubators in Philip Cooke *et al*.'s contribution entitled 'The Biosciences Knowledge Value Chain and Comparative Incubation Models.' These authors emphasize the ways in which regional capabilities condition the scale of operations necessary for such interactions to work. They contrast the boundary-crossing issues for biotechnology spinout incubation between regions such as Sardinia with a genetic resource (the 400-year-old family records of an insular population) but with little research or knowledge exploitation capability and, among numerous others, Massachusetts, where $1 billion in public health-care research funding is spent each year on 'ahead of the curve' genomic research. This has attracted acquisitions from the likes of Pfizer, Abbott, Wyeth, Amgen, and AstraZeneca and a $250 million Institute of Genomics by Novartis, and shows how the local Cambridge biosciences cluster has spawned a *constructed advantage* statewide and for the U.S. by its magnetizing effects upon firms, policies and talent. Bioincubation, even in distant Worcester, Massachusetts, is a 'no-brainer' provided spinouts arrive with the three key assets of a business plan, IPR, and finance – as they invariably do, 'friends, family, and fools' being the principal financiers.

**Knowledge and Regionally Constructed Advantage**

So what is the difference between a knowledge-based economy and a knowledge economy? For Dunning (2000) they are the same since his book title refers to the former while his introductory chapter refers to the latter. Two of its 'key engines' are 'the microchip and the computer' (p. 9); yet these are pervasive across sectors, but the key knowledge is technological. For Machlup, as we have seen, the knowledge economy is a set of sectors which intensely concentrate knowledge assets in terms of both human and fixed capital. This remains the kind of measure favoured by international economic organizations like the OECD and EU, as we have seen. Does this mean that actors not included in Machlup's six knowledge sectors are robots without knowledge?

Let us distinguish among knowledge of the *analytical* (science), *synthetic* (technical) or *symbolic* (creative) kind. In all sectors, knowledge has become significantly more important than in previous configurations (industry-based or agrarian economies). So we may conclude that as the *base* of knowledge evolves institutionally, an increasing portion of the economy becomes knowledge-intensive. One key difference, however, is that science-based industries like genomics, research, software and nanotechnologies generate value from producing analytical knowledge while most others create value from exploiting synthetic or symbolic knowledge. Thus, the old definition of *knowledge economy* in terms of a few important and growing sectors is redundant, while the structural idea of a *knowledge-based economy* linking the knowledge generation sub-system (mainly laboratory research) to the knowledge-exploitation system (mainly firms and, say, hospitals or schools) via technology transfer organizations in regional innovation systems is analytically useful.

The effect of the growth in importance of regional (and other) innovation systems is to pervade the regional and other economies with scientific, synthetic and symbolic knowledge to a greater extent than ever before. The organization of pure and applied knowledge can increasingly pervade the economy when scientific and technological knowledge is institutionally produced and systematically controlled. R&D management and S&T policies at relevant government levels enlarge the set of options. These, however, are not fixed but evolving distributions in which some regions are more developed as knowledge-based economies than others. Hence, the post-1970s fascination with 'high-tech' regions worldwide. Today, however, as the Triple Helix perspective suggests, with universities and their related research laboratories spread throughout most regions, many more economies have the chance to access not only yesterday's 'global' knowledge announced on the Internet and exploitable by all, but local knowledge of potentially high value generated from research conducted in relation to regional capabilities. Thus, as the knowledge base becomes pervasive, the knowledge economy is further reinforced.

The paper in this collection by Gary Gorman and Sean McCarthy entitled 'Business Development Support and Knowledge-Based Businesses' addresses this issue in terms of the knowledge requirements of business. In 'Business Development Capabilities in Information Technology SMEs in a Regional Economy: an exploratory study' Charles Davis and Elaine Sun focus on a specific sector within this same domain. Both papers explore the problems of business development in a localized region like Atlantic Canada.

The constructed advantage that may accrue from innovation systems designed in relation to regional capabilities is examined in the papers by Bjørn Asheim & Lars Coenen and by Janet Bercovits & Maryann Feldman. In 'Contextualizing Regional Innovation Systems in a Globalising Learning Economy: On Knowledge Bases and Institutional Frameworks' Bjørn Asheim & Lars Coenen give special importance to the linkage between the larger institutional frameworks of the national innovation and business systems and the character of regional innovation systems. In the paper by Janet Bercovitz & Maryann Feldman entitled 'Entrepreneurial Universities and Technology Transfer: A Conceptual Framework for Understanding Knowledge-based Economic Development,' the Triple Helix challenge is picked up in an attempt to identify the factors that affect the ability of universities both to create new knowledge and to deploy that knowledge in economically useful ways and thereby contribute to economic growth and prosperity.

It seems therefore that *constructed advantage* based on regional innovation systems that *transceive* over long distances as well as through regional networks is becoming the model of choice for achieving accomplished regional economic development. The importance of effective communication in this process is highlighted in Loet Leydesdorff's contribution to the issue entitled '"While a Storm is Raging on the Open Sea": Regional Development in a Knowledge-based Economy.' Leydesdorff argues that the knowledge base of an economy can be considered as a second-order interaction effect among Triple Helix interfaces between institutions and functions in different spheres. Proximity enhances the chances for couplings and, therefore, the formation of technological trajectories. In this manner, connections between regional innovation systems and markets (an understudied aspect in the broad field of innovation studies) may be facilitated.

**Conclusion**

Our own contributions and the one by Bercovitz & Feldman, which also focuses on boundary-crossing problems experienced by universities in relation to markets, examine how small events triggered in specific institutions, often in proximity, can exert a global impact in fields like the treatment of hitherto incurable diseases by new biotechnologically derived treatments. The transformation is focused: a trajectory can be shaped at some places, but not at others. A lock-in functions like a resonance which transforms the resonating dynamics. It cannot be known *ex ante* which dimensions in the multi-dimensional arrangement of industry, academia, and governance will be able to retain wealth from the incursive transformation. However knowledge-intensive, the geographical dimension remains always involved because the events are also localized (Storper, 1997).

Geographical proximity can be expected to serve the incubation of new technologies. However, the regions of origin do not necessarily coincide with the contexts that profit from these technologies at a later stage of development. The dynamics can evolve with the technology (Hughes, 1987). Various Italian industrial districts, for example, have witnessed a flux of new developments. As companies develop a competitive edge, (some of) their activities may move out of the region generating a threat of deindustrialization which has to be countered continuously at the regional level (Dei Ottati, 2003; Sforzi, 2003). The four regions indicated by the EU as 'engines of innovation' in the early 1990s were no longer the most innovative regions in the late 1990s (Krauss & Wolff, 2002; Laafia, 1999; Viale & Campodall'Orto, 2002).

The technological trajectory serves as a pathway for a next-order regime to become established. Although the regime can be considered as operating like an attractor from the perspective of hindsight, the technological landscape is yet a *terra incognita* for the actors involved. They operate in a concrete landscape, but with reflexive expectations. The reflections enable the agents to explore new options. Schumpeter (1943) called this 'creative destruction.' The dimensions (subdynamics) that prove to be most important for realizing the new options can only be determined *ex post.* The 'lock-ins' leading to growth can be expected to happen for stochastic reasons (Arthur, 1994). The spatial perspective is only one among various possible perspectives on the innovation system (Bathelt, 2003). One can also assume a sectoral or a technological perspective, but this leads to a different research design (Leydesdorff, 2001). The interactions between technologies and economies, however, remain constrained and enabled by the historical contingencies of the system; the localized perspective provides us with a focus on the retention mechanism.

---

*Table of Contents* of **The Knowledge-Based Economy and Regional Development**, special issue of the *Journal of Technology Transfer*. Philip Cooke & Loet Leydesdorff (eds.)

- Janet Bercovitz & Maryann Feldman, 'Entrepreneurial Universities and Technology Transfer: A Conceptual Framework for Understanding Knowledge-based Economic Development.'

- Loet Leydesdorff, '"While a Storm is Raging on the Open Sea": Regional Development in a Knowledge-based Economy.'